\documentclass[11pt,a4paper]{article}
\usepackage[margin=25mm]{geometry}
\usepackage{amsmath,amssymb,graphicx,booktabs}
\usepackage[authoryear,round]{natbib}
\numberwithin{equation}{section}
\makeatletter

\newenvironment{fixedfigure}{\par\addvspace{\intextsep}\noindent
 \begin{minipage}{\linewidth}\def\@captype{figure}\centering}
 {\end{minipage}\par\addvspace{\intextsep}}
\makeatother
\usepackage[hidelinks]{hyperref}
\hypersetup{pdftitle={Is the Quiet-Sun Heating Requirement Observationally Established?},pdfauthor={Victor Edmonds}}

\graphicspath{{figures/}}
\newcommand{\Te}{T_{\mathrm e}}
\newcommand{\Tr}{T_{\mathrm r}}
\newcommand{\Tc}{T_{\mathrm c}}
\newcommand{\Thyd}{T_{\mathrm H}}
\newcommand{\kb}{k_{\mathrm B}}
\newcommand{\me}{m_{\mathrm e}}
\newcommand{\nelec}{n_{\mathrm e}}
\newcommand{\Dkl}{D}
\title{Is the Quiet-Sun Heating Requirement Observationally Established?}
\author{Victor Edmonds\textsuperscript{1,2}\\[0.5em]
\small \textsuperscript{1}Final Stop Consulting LLC, Fuquay-Varina, NC, USA\\
\small \textsuperscript{2}Ronin Institute for Independent Scholarship 2.0, Sacramento, CA, USA\\[0.3em]
\small \href{mailto:vedmonds@finalstopconsulting.com}{vedmonds@finalstopconsulting.com}}
\date{}

\begin{document}
\maketitle

\begin{abstract}
How much energy does the quiet corona require? Its radiation shows
that energy must be supplied. Electrons also carry heat into or out of
it. We ask whether temperature observations justify the heat-flow
estimate used to judge a heating mechanism's energy supply.
Two radio methods give temperatures near 0.62 and 1.5
million kelvin. A population with extra fast electrons can give both
readings. Unresolved structure and delayed ion formation also leave
Maxwellian explanations open. Neither temperature interpretation alone
determines the heat flow. The tested local conduction formula diverges
for the idealized fast-electron model with an unbroken tail. Applying
X-ray limits to the model still leaves about 5--6\% of the electrons
holding 28--33\% of its kinetic energy. The heat they carry depends
on the balance between directions and on conditions along their paths.
Source and collision calculations test whether the population could
be sustained. The observations examined here do not establish whether
the conductive estimate is accurate enough for the heating comparison.
We cannot judge whether a mechanism supplies enough energy against
a heat loss we have not established. Measuring, reliably calculating,
or sufficiently bounding that loss would settle this part of the budget.
\end{abstract}


\section{How much energy does the quiet corona require?}\label{sec:budget}

How much energy must a heating mechanism supply to the quiet corona?
The corona radiates energy, and electrons carry heat across its
boundaries. Radiation can be observed. The electron heat flow must
be measured, reliably calculated, or bounded.

We cannot judge whether a heating mechanism supplies enough energy
against a heat loss we have not established. The estimate could be right. The question is whether the evidence
establishes that it is accurate enough for the heating comparison. A measurement or a sufficiently
restrictive bound on heat flow could meet it without determining
the full electron distribution.

Two radio methods give temperatures near 0.62 and 1.5 million kelvin
\citep{mercier2015,mercier2016corr}. Earlier work examined whether one electron
population could give both readings \citep{edmonds2026a}.
Subsequent calculations examined why spectroscopy can return familiar
temperatures and why a conventional differential-emission-measure
inversion can miss electron-distribution structure
\citep{edmonds2026conv,edmonds2026dem}. Here, we follow the
consequences for heat transport. Which information about the electrons
is needed to calculate the heat they carry?

\begin{figure}
 \centering
 \includegraphics[width=\linewidth]{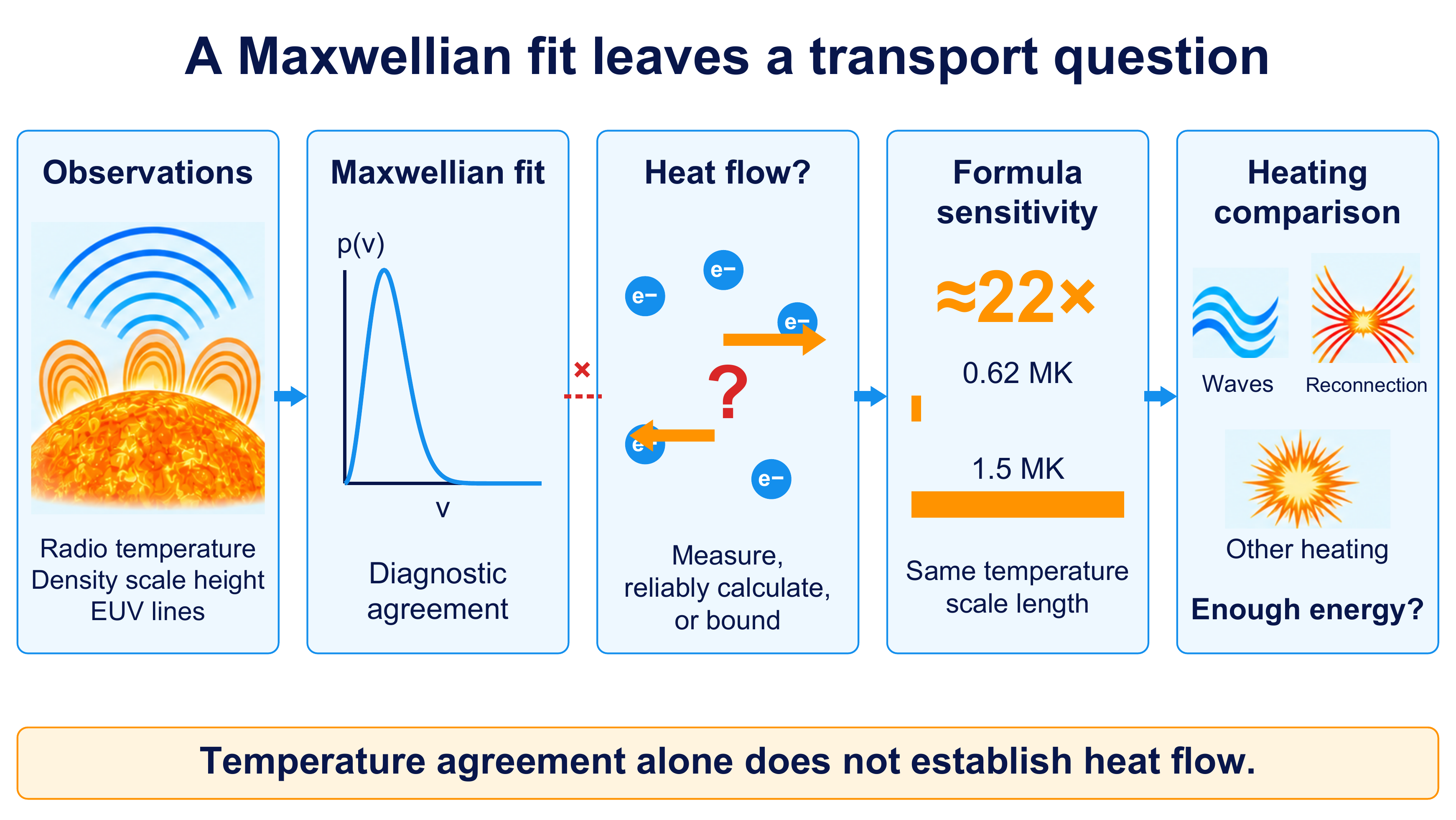}
 \caption{A successful Maxwellian temperature fit still leaves the
 heat-flow calculation to be justified. Its collision rates, gradients,
 and geometry require separate checks. The bars show the approximately
 22-fold change in the
 classical heat-flux estimate between 0.62 and 1.5 MK at fixed
 temperature scale length and coefficient (equation~\ref{eq:SH}).
 This is a formula sensitivity, not a measured flux ratio or a ratio
 of total heating requirements. Curves and icons are schematic. Decorative illustrations were
 generated with OpenAI image-generation assistance; text and quantitative
 graphics were drawn with a plotting script.}
 \label{fig:maxwellian-inference}
\end{figure}

We test a population with extra fast electrons and examine Maxwellian
alternatives. These are possible explanations of the temperature
readings. Each interpretation must support a calculation of heat flow
before that calculation can establish the conductive energy demand.
Figure~\ref{fig:maxwellian-inference} shows the additional transport
requirements even when a Maxwellian diagnostic interpretation is granted.

Classical conduction along the magnetic field has the form
\begin{equation}\label{eq:SH}
 q_{\parallel,\mathrm{SH}}=-\kappa_0 T^{5/2}\frac{\partial T}{\partial s},
\end{equation}
where $s$ measures distance along the magnetic field and $\kappa_0$
contains the collisional coefficient. This expression is derived for
electrons close to a local Maxwellian, with collisions frequent enough
and gradients gentle enough for the local approximation
\citep{spitzer1953,braginskii1965}. At fixed
gradient length $L_T=T/|\partial T/\partial s|$ and fixed coefficient,
its magnitude scales as $T^{7/2}$. Substituting $1.5$ rather than
$0.62$~MK changes the result by a factor of about 22. The rounded pair
$1.5$ and $0.6$~MK gives about 25. This shows the formula's sensitivity
to temperature; it does not measure an error in the Sun's heat flux.
We must establish what the temperature represents before using it
in the energy balance.

The energy balance also depends on the volume chosen. For a steady
coronal volume with negligible bulk flow, it has the schematic form
\begin{equation}\label{eq:balance}
 \int_V Q\,dV=\int_V L_{\mathrm{rad}}\,dV+
 \oint_{\partial V}\mathbf q\cdot d\mathbf A,
\end{equation}
where $Q$ is the net volumetric energy input, $L_{\mathrm{rad}}$ is the
radiative loss rate and $\mathbf q$ is the conductive heat flux.
The sign of the boundary term distinguishes energy entering from energy
leaving. Flows, changes in stored energy, and other energy carriers add
their own terms \citep{withbroe1977}. Heat flowing downward can be
a loss from the coronal volume and an energy supply to the transition
region below it. It is counted once in the balance of the combined volume.

Coronal-heating models use this energy balance to test how different
mechanisms could maintain the observed corona
\citep{cranmerwinebarger2019}. In a recent forward-model comparison,
\citet{cranmergilly2026} adopt a conductive prescription, calculate
coronal structure for different heating prescriptions, and compare
the resulting extreme-ultraviolet (EUV) and X-ray emission with
observations. Agreement
supports the combined model. The question pursued here is what
evidence establishes the conductive contribution accurately enough
to use it in judging the required heating.

Kinetic studies have long examined how electron distributions and
transport affect coronal energy balance
\citep{shoub1983,scudder1992a,landi2001}. The recent review by
\citet{coburn2026} describes continuing difficulties in connecting
collisional, collisionless, and turbulent electron transport.
Here, we connect that transport question to a specific observational
problem: do the quiet-Sun temperature measurements constrain the
electrons well enough to justify the heat flow calculated from them?

The transport assumption enters published heating estimates.
\citet{maccormack2017} combine EUV images with
a magnetic-field model to estimate radiation and heat flow through the boundaries of
quiet-Sun coronal loops. They calculate the heat flow using
equation~\ref{eq:SH}, then compare the required energy supply with
estimates of energy carried by Alfv\'en waves. Their typical conductive
term is smaller than the radiative term and can represent a gain or
a loss. The question raised here concerns how that conductive term
is justified, and comparisons that depend on it. The calculations
below test the step from temperature to transport; they do not recalculate
those loop budgets.

Radiation from the transition region offers another way to constrain
energy crossing a coronal boundary. It can constrain downward heat
flow if local heating, energy carried by bulk flows, changes in stored
energy, and the other boundary terms are independently small or bounded.
\citet{fontenla1991} compare transition-region radiation with downward
particle heat flow in hydrostatic models. Such an energy-balance
constraint does not have to recover the full electron distribution or validate
a local conduction law. The temperature comparisons studied here do
not test whether this route gives a sufficiently accurate conductive
constraint.

Sections 2--3 examine what the temperature measurements constrain and
test alternative electron populations. Section 4 asks what else is
needed to calculate conduction. Section 5 tests the fast population's
high-energy extent, maintenance, and source. Those tests ask whether
the population could exist. A measured energy flux could establish
part of the budget before its microscopic source was known.

\section{Why do the temperature readings differ?}\label{sec:diagnostics}

\subsection{What a brightness and a scale height measure}\label{sec:thermometers}

Brightness temperature is a convenient way to express specific
intensity. In the Rayleigh--Jeans limit,
$I_\nu=2\kb\nu^2T_b/c^2$. For a homogeneous source with negligible
background, radiative transfer gives
$T_b=T_S(1-e^{-\tau_\nu})$, where $T_S$ expresses the source function
as a temperature. The optically thick limit reads $T_S$; an optically
thin brightness also depends on optical depth. We denote the fitted
coronal radio source temperature by $\Tr$ to distinguish it from a
brightness measured in an arbitrary image pixel.

\citet{mercier2015} used six frequencies between 150 and 450~MHz on
183 quiet days during 2004--2011. Their radio source and scale-height
temperatures motivate the ratio $R_H=\Thyd/\Tr\simeq2.4\pm0.3$
used here and in \citet{edmonds2026a}. These are inferred physical
quantities obtained from intensity and spatial structure, rather than
two direct measurements of the same electron moment. Their different
spatial weightings are part of the problem. Mercier and Chambe
interpreted the difference in terms of hotter ions.

The scale-height interpretation begins with pressure balance. For an
isotropic electron population, the pressure defines a mean-energy
temperature through $P_e=\nelec\kb\Te$ even when the distribution
is not Maxwellian. Hydrostatic balance, however, involves total
pressure. For fully ionized hydrogen with helium abundance
$n_{\mathrm{He}}/n_{\mathrm H}=0.1$, a common-temperature hydrostatic
fit corresponds, in the simple common-profile approximation, to
\begin{equation}\label{eq:ions}
 \Thyd=\frac{\Te+0.917T_i}{1.917},\qquad
 \Te=1.917\Thyd-0.917T_i,
\end{equation}
where $T_i$ is the ion-number-weighted ion temperature.
Equal ion and electron temperatures give $\Te=\Thyd$. A spatially
varying atmosphere requires the full pressure-gradient equation;
flows, geometry, and unequal temperature profiles cannot in general be
represented by one algebraic correction. Equation~\ref{eq:ions}
therefore states an assumption behind the electron interpretation of
the fitted scale height, not an independent determination of $\Te$.

\subsection{A distribution that gives two different readings}\label{sec:kappa}

Let $f(\mathbf v)$ be the electron velocity probability density in
the electron rest frame, normalized to unity. Its mean-energy
temperature is defined by
\begin{equation}\label{eq:Te}
 \frac32\kb\Te=\int\frac12\me v^2 f(\mathbf v)\,d^3v.
\end{equation}
A useful test family is the isotropic $\kappa$ distribution
\begin{equation}\label{eq:kappa}
 f_\kappa(v)=
 \frac{\Gamma(\kappa+1)}{\Gamma(\kappa-1/2)(\pi\kappa v_c^2)^{3/2}}
 \left(1+\frac{v^2}{\kappa v_c^2}\right)^{-\kappa-1},
 \qquad v_c^2=\frac{2\kb\Tc}{\me},
\end{equation}
with $\kappa>3/2$ for a finite mean energy
\citep{vasyliunas1968,dzifcakova2013}. Here, $\Tc$ is a scale
parameter defined by this equation. In particular, it should not be
silently equated with the temperature of every possible fit to the
low-energy core. For the unbroken family,
\begin{equation}\label{eq:temperatures}
 \Te=\frac{\kappa}{\kappa-3/2}\Tc.
\end{equation}

The low-frequency free-free source function provides a physical reason
to compare this family with the radio discrepancy. For an isotropic
$\kappa$ distribution, the slowly varying Gaunt-factor approximation
in \citet[][equation~48]{fk2014} gives $\Tr\simeq\Tc$.
The source function and the pressure thus weight the distribution
differently. Combining that approximation with $\Te=\Thyd$ gives
\begin{equation}\label{eq:inversion}
 R_H\simeq\frac{\kappa}{\kappa-3/2},\qquad
 \kappa\simeq\frac{3R_H}{2(R_H-1)}.
\end{equation}
The central ratio gives $\kappa=2.57$; the adopted range alone maps
to approximately $2.38$--$2.86$. Ion-pressure assignments, radio
transfer, and geometric systematics add model dependence. The
representative $\kappa=2.5$ example used below has an unbroken ratio
of 2.5, which is distinct from the central observed ratio 2.4.

The value of this construction is that it makes a proposed explanation
calculable. It specifies what a cool radio reading and a larger
pressure temperature could mean for one population. It does not
establish that the Sun has chosen this family, or that the fitted
scale height has isolated the electron pressure. An alternative family
requires its own radio response, and a structured Maxwellian atmosphere
requires its own transfer calculation. The discrepancy survives as an
observational question when a particular interpretation fails.

\subsection{Why familiar spectroscopic temperatures can still appear}\label{sec:spectroscopy}

An electron-impact rate is an integral of the distribution against an
energy-dependent cross-section. Schematically,
\begin{equation}\label{eq:rate}
 C_j[f]=\int_{E_{\mathrm{th},j}}^\infty
 \sigma_j(E)v(E)p(E)\,dE,
\end{equation}
where $p(E)$ is the normalized energy distribution and
$E_{\mathrm{th},j}$ is the threshold. A Maxwellian temperature fitted
to such a rate is the temperature that reproduces that response.
Different cross-sections do not necessarily return the same fitted temperature
for a non-Maxwellian population. Ion fractions add the competition
between ionization and recombination; line intensities add excitation,
density, and integration through the source
\citep{owockiscudder1983,dzifcakova2013,dudik2014}.

This explains how spectroscopic agreement with a Maxwellian can coexist
with unresolved distribution structure. The earlier calculations in
\citet{edmonds2026conv,edmonds2026dem} examined that loss of
discrimination. In the particular Fe\,\textsc{xii}/Fe\,\textsc{xiii}
ionization-table example examined in the underlying analysis, a
$\kappa=2.5$ population at mean temperature $1.5$~MK returns a
Maxwellian-equivalent temperature near $1.47$~MK. This is a worked
response example, not a proof that all spectroscopic temperatures
equal the mean energy or that a complete observed spectrum is
reproduced. The rates, line set, and formation weighting must be tested
for the application in question \citep{dz23}.

\citet{lorincik2020} used
diagnostics explicitly sensitive to $\kappa$, together with density
and temperature-distribution modelling, and found their quiet-Sun
spectra consistent with a Maxwellian, in contrast to their active-region
results. A proposed quiet-Sun tail must account for that result.
The worked response above does not reproduce their multi-line analysis.
A Maxwellian-consistent spectrum also does not, by itself, determine
how much heat electrons carry. Its temperature and structure must be
connected to heat flow by a transport model. We next test what an
additional line-ratio comparison and a Maxwellian reconstruction can
establish about the atmosphere.

\section{Can a Maxwellian account for the observations together?}\label{sec:alternatives}

A discrepancy is useful only if plausible ways of producing it are
examined. The relevant alternatives involve different physics:
calibration changes a measured line ratio, temperature structure changes
the mixture of emitting plasma, and delayed ionization changes which
ions are present at a given electron temperature. Unresolved structure
also changes how radio brightness represents the atmosphere. None of
these possibilities can be settled by choosing a distribution family.

\subsection{An additional spectroscopic test}\label{sec:feix}

Lines of the same ion offer a useful additional test because their
ratios do not depend directly on the abundance of that ion in a single
uniform source. We use Fe\,\textsc{ix} lines observed by the Extreme
Ultraviolet Imaging Spectrometer (EIS) on \textit{Hinode}. The principal
archival example is the 11 March 2007 quiet-Sun raster used in the
benchmarks of \citet{delzanna2010,delzanna2012}, with a bright
atlas region and an inner off-limb annulus analysed separately. The
atomic calculation uses the 915-level Fe\,\textsc{ix} model of
\citet{dzsbm2014}. Appendix~\ref{app:feix} gives the test definitions,
calibration checks, and limits of the additional population comparisons.

The experiment needs a check on its own calibration. The ratio of
the 189.941 and 177.592~\AA\ lines is comparatively insensitive to
the distribution in the tested models, while the 197.862~\AA\ line
provides additional sensitivity. The first ratio therefore checks
whether an apparent discrepancy involving the second can safely be
assigned to the electrons. With the baseline calibration, the inner
annulus fails this check. With the alternative
\citet{delzanna2013cal} calibration, the check passes, but the primary
distribution test does not cross its specified rejection threshold.
The absolute test consequently does not establish a non-Maxwellian
distribution.

There is nevertheless a useful residual comparison. In photon units,
the nearby-line ratio
\begin{equation}\label{eq:r197}
 r_{197}=I(197.862~\text{\AA})/I(189.941~\text{\AA})
\end{equation}
is $1.291$ in the atlas region and $1.362$ in the inner annulus under
the baseline calibration. A particular narrow-temperature Maxwellian
model gives approximately $0.989$; the adopted broad
differential-emission-measure model gives $1.132$. A differential
emission measure specifies how much emitting material contributes at
each temperature. Broadening it thus changes a line ratio even when
every local population is Maxwellian. With the inherited 9\% systematic
allowance, the narrow Maxwellian differs by about 2.8 allowance units
in each region, whereas the broad model differs by approximately
1--2.3. These units are a way of reporting a residual against an
adopted uncertainty allowance, not Gaussian detection significances.
Under the narrow-temperature treatment, the $\kappa=2.5$ and 3
examples have smaller residuals than the Maxwellian. That preference
is worth retaining, together with the broad-Maxwellian alternative
that weakens it. Appendix~\ref{app:feix} reports the treatment and
density checks. The line measurements constrain a comparison among
specified models; they do not select a unique distribution.

Calibration also changes the comparison itself: the 2013 calibration
reduces the two observed ratios to $1.198$ and $1.263$. Furthermore,
the Fe\,\textsc{ix} 189.94/197.85~\AA\ pair enters the constraints used
in the updated EIS calibration \citep{dzww2025}. A subsequent use of
that calibrated pair must account for this dependence before treating
agreement as an independent test of the assumed atomic or distribution
model. This is a requirement on the inference, not a criticism of
using physical line constraints to calibrate an instrument.

\subsection{Could time-dependent ion formation produce the ratio?}\label{sec:nei}

Electrons can change temperature before ions complete their adjustment
to a new ionization balance. An ion may then emit at temperatures
where an equilibrium calculation would predict very little of it.
This offers a Maxwellian alternative to interpreting a line ratio as
evidence for a suprathermal electron population. Time-dependent
diagnostics and cooling observations motivate taking the possibility
seriously \citep{doyle2012,viall2012}; those studies do not themselves
provide a model for this quiet-Sun raster.

We make a bounded test: can Maxwellian excitation reach the observed
ratio if the amount of Fe\,\textsc{ix} is temporarily left unconstrained?
The calculation solves the level populations over
$5.3\leq\log_{10}(T/\mathrm K)\leq6.8$ and
$7.8\leq\log_{10}(\nelec/\mathrm{cm}^{-3})\leq9.0$, retaining
statistical equilibrium among the excited levels but removing
equilibrium ion-fraction weighting. At
$\nelec=2\times10^8$~cm$^{-3}$, the baseline atlas and annulus ratios
are reached near 2.07 and 2.62~MK. With the 2013 calibration, the
crossings occur near 1.55 and 1.89~MK. Maxwellian excitation therefore
cannot be excluded by these ratios alone.

Attaining a ratio is less demanding than producing its measured
intensities. At the two baseline crossings, the equilibrium
Fe\,\textsc{ix} fractions are only about $4.1\times10^{-4}$ and
$4.0\times10^{-6}$. A viable history would have to preserve enough
Fe\,\textsc{ix}, for long enough, while reproducing the other lines.
No such history is calculated here. The test also omits time-dependent
excited-level populations and recombination feeding. Its result is
specific: the observed ratios lie within a Maxwellian excitation
envelope, so the equilibrium ratio comparison is not a general
exclusion of time-dependent Maxwellian plasma.

\subsection{Could radio structure account for the lower temperature?}\label{sec:radio}

For a uniform Maxwellian layer, the radio brightness temperature is
\begin{equation}\label{eq:slab}
 T_b(\nu)=T_{\mathrm{bg}}e^{-\tau_\nu}
       +T(1-e^{-\tau_\nu}),
\end{equation}
where $\tau_\nu$ is the free-free optical depth and $T_{\mathrm{bg}}$
is the background brightness. In a structured atmosphere, transfer
along a ray and averaging across the telescope beam perform different
operations. A thick component behind another thick component can be
hidden. A thick component occupying only part of the beam can instead
be diluted by the remaining area. The latter matters for the
approximately 100-arcsec resolution relevant to the radio comparison.
The transfer uses the free-free prescription of \citet{dulk1985}.
Propagation adds further complications at metre wavelengths
\citep{sharma2020}.

A line-of-sight experiment used a prescribed emission
measure distribution with total column emission measure
$2.44\times10^{27}$~cm$^{-5}$ and emission-measure-weighted temperature
$T_{\mathrm{EM}}=1.44$~MK. Putting its coolest material in front
gave a fitted radio temperature near $1.19$~MK. This corresponds to
$T_{\mathrm{EM}}/T_{\mathrm{fit}}\simeq1.21$. It is a useful result
for that atmosphere, but $T_{\mathrm{EM}}/T_{\mathrm{fit}}$ is not
the observed hydrostatic-to-radio ratio $\Thyd/\Tr$. The calculation
tests this reconstruction rather than establishing a general upper
bound on the observed ratio. Ordering the parcels also bounds the
brightness at each frequency; it does not by itself bound the
temperature returned by a nonlinear spectral fit.

\begin{figure}
 \centering
 \includegraphics[width=\linewidth]{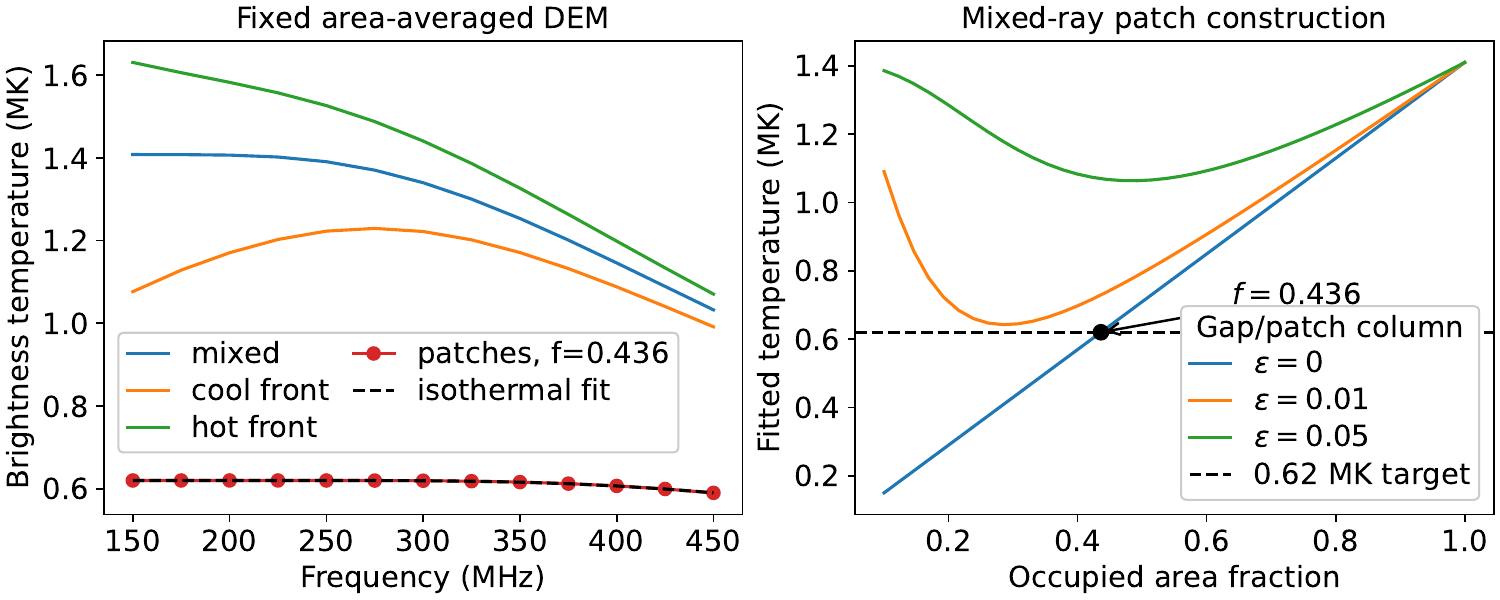}
 \caption{Maxwellian radio transfer with a fixed area-averaged
 emission measure distribution. Left: different arrangements along
 a ray and a transverse patch model. The patch spectrum closely
 resembles a cooler isothermal spectrum. Right: the mixed-ray patch
 construction as occupied area and gap emission change; the marked
 point gives a fitted temperature of 0.62 MK. This demonstrates an
 ambiguity in fitting brightness temperatures. The calculation does
 not reproduce the observed spatial scale height, scattering,
 refraction, or joint radio and EUV data. The figure was generated
 with a Python plotting script prepared with Codex (OpenAI) assistance.}
 \label{fig:radio}
\end{figure}

Allowing transverse structure changes the outcome. In an illustrative
patch model, the same area-averaged emission measure is concentrated
into a fraction $f$ of the beam, with negligible coronal emission in
the remaining area. An occupied fraction $f\simeq0.436$ produces
$T_{\mathrm{fit}}=0.62$~MK, with a maximum fractional residual of
about $0.04\%$ from an isothermal free-free fit at the 13 synthetic
frequencies between 150 and 450~MHz (figure~\ref{fig:radio}). These
are synthetic frequencies, not a new fit to the six-frequency NRH data.
Under the density convention of this model, the occupied column would
require a path of about 1400~Mm at $2\times10^8$~cm$^{-3}$, or a
density near $1.1\times10^9$~cm$^{-3}$ for a 50~Mm path. Such choices
would have to remain compatible with the spectroscopic density and
the geometry.

The construction is sensitive to how empty the gaps are. Let
$\epsilon$ be the gap column emission measure divided by the
occupied column emission measure. Holding the area average fixed
then gives
\begin{equation}\label{eq:patchcolumns}
 \mathrm{EM}_{\mathrm{occ}}=
 \frac{\overline{\mathrm{EM}}}{f+(1-f)\epsilon},\qquad
 \mathrm{EM}_{\mathrm{gap}}=\epsilon\mathrm{EM}_{\mathrm{occ}}.
\end{equation}
For the mixed-ray construction shown in figure~\ref{fig:radio},
setting $\epsilon=0.01$ or 0.05 raises the minimum fitted
temperature in the tested grid to approximately 0.64 or 1.06~MK.
Other arrangements give different minima. A constant filling fraction can lower an inferred density
normalization without changing an optically thin density scale height;
a fraction that varies with height can also change the slope. Neither
behaviour has been inferred from resolved images in this calculation.
The result is enough to keep unresolved structure in contention,
but not enough to identify it as the observed explanation.

The joint question therefore remains: can unresolved Maxwellian
structure reproduce the radio temperature while also satisfying the
density, spatial, and spectroscopic constraints? The patch calculation
answers the spectral part. Its column requirements expose what the
rest of the explanation must supply. The time-dependent Fe\,\textsc{ix}
alternative likewise reaches a ratio without yet reproducing the
intensities. Both prevent a general exclusion of Maxwellian plasma;
neither closes the joint observational problem. The conditional
$\kappa$ interpretation has not supplied that reconstruction either,
and the radio and EUV records are not co-spatial or contemporaneous.

Even if one atmosphere reproduces all these observations, its heat
loss by conduction must still be established. The needed calculation
depends on how electrons respond to temperature gradients and
boundaries, as well as on the temperatures inferred from radiation.

\section{What changes if the electrons have a fast tail?}\label{sec:transport}

\subsection{A temperature does not determine a heat flux}\label{sec:heatflux}

Heat flux depends on how much energy electrons carry in opposing
directions. Equal contributions cancel even when the electrons
contain substantial kinetic energy. This directional balance appears
explicitly in the heat-flux integral.
In the electron rest frame, for a distribution normalized to unity,
\begin{equation}\label{eq:heatflux}
 \mathbf q=\nelec\int \frac{\me c^2}{2}\,\mathbf c\,
                     f(\mathbf c)\,d^3c,
\end{equation}
where $\mathbf c$ is velocity relative to the bulk electron flow.
An isotropic distribution has zero heat flux, even if it has a large
suprathermal tail. The outward and inward contributions cancel.
The isotropic models used above can therefore constrain energy
content and diagnostic responses, but do not supply a measured
conductive flux.

A transport closure supplies the additional relation between such
moments and the lower-order quantities that a fluid model evolves.
In classical conduction, a nearly Maxwellian population and a small
collisional departure from it connect heat flux to the local temperature
gradient. The relevant collisional ordering must hold for the electrons
that carry the energy, not just for those near the centre of the
distribution. In the fast-electron Coulomb limit, the collision time
scales approximately as $v^3$ and the travel distance between
collisions as $v^4$. An electron at three times a reference speed can
therefore sample a distance approximately 81 times larger in this
scaling. A local thermal mean free path is not, by itself, a check
on the locality of the energetic contribution
\citep{scudderolbert1979,shoub1983}.
The distribution underlying an approximate closure must also remain
physically admissible. \citet{scudder2021} evaluates how negative
phase-space densities in truncated Spitzer and Grad expansions can
affect the heat-flux moment itself.

\subsection{A concrete limit of a local conductive response}\label{sec:convergence}

The distinction between a finite temperature and a defined conductive
response can be made explicit. In the Lorentz-plasma approximation used here, electron--ion
collisions are represented by a speed-dependent Krook relaxation
rate proportional to $v^{-3}$.
The local linear response about an unbroken $\kappa$ distribution
contains an integral proportional to
\begin{equation}\label{eq:responseintegral}
 I_\kappa=\int_0^\infty
 v^{11}(1+Av^2)^{-\kappa-2}\,dv,
 \qquad A=(\kappa v_c^2)^{-1},
\end{equation}
as in \citet{du2013} and the corrected derivation of
\citet[][equations~18 and 35, Appendix~B]{husidic2021}. At large speed its integrand scales as
$v^{7-2\kappa}$, so convergence requires $\kappa>4$.
Truncating this integral at a large speed $V$ gives growth
proportional to $V^3$ at $\kappa=2.5$, $V^2$ at $\kappa=3$,
and $\ln(V/v_c)$ at $\kappa=4$. Continuing a closed-form expression
beyond its integral's convergence domain does not define the missing
conductivity. The pressure can be finite while this particular
conductive response is not.

This is a failure of the specified unbroken local response, not a
nonzero heat flux of the isotropic reference or a theorem excluding
every fluid description. Generalized moment closures exist
\citep{levermore1996,cranmerschiff2021}. Regularizing the tail also
restores finite coefficients, with values that depend on the
regularization \citep{husidic2022}. For the terminated examples
below, the unbroken divergence cannot simply be carried over:
the continuation and its response must be included consistently.

For this local response, knowing the temperature and $\kappa$ does
not give a usable coefficient in the regime examined. A cutoff removes
the ideal-tail divergence, but a finite coefficient alone does not
show that a local law applies. The calculation must still describe
how electrons respond to the gradients and boundaries. That connection
has not been established for the candidate atmosphere here, so the
conductive loss for its energy budget remains to be calculated.

\subsection{How much energy can a small population contain?}\label{sec:tail}

The $\kappa=2.5$ reference makes the stakes tangible. Above
$v=2.1\sqrt{2\kb\Te/\me}$, approximately 6\% of the electrons
carry 40\% of its kinetic energy. Here, $\Te$ is the mean-energy
temperature of the unbroken distribution. The fractions are integrals
of the specified model, not direct counts in the quiet corona. They
show why a response dominated by more numerous slower electrons does
not necessarily settle the energy content of the faster population. Conversely,
a mean-energy temperature counts that energy without identifying
which electrons carry it.

An unbroken power law is not a realistic unrestricted extrapolation.
Section~\ref{sec:xray} applies hard-X-ray limits to explicitly
terminated versions of the same family. At the endpoints of the
allowed sharp-cutoff example, about 5.1--5.7\% of the electrons carry
28--33\% of the kinetic energy above the corresponding reference
threshold. The tail remains energetically consequential even after
the high-energy restriction: the energy concentration is not solely
an artifact of extending a power law to unlimited energy. Its contribution to heat flow, however,
still depends on the imbalance between directions and on how the
population responds to its surroundings.

Figure~\ref{fig:kappa-inference} follows this conditional interpretation
from the diagnostics to its energetic and transport obligations.

\begin{fixedfigure}
 \centering
 \includegraphics[width=\linewidth]{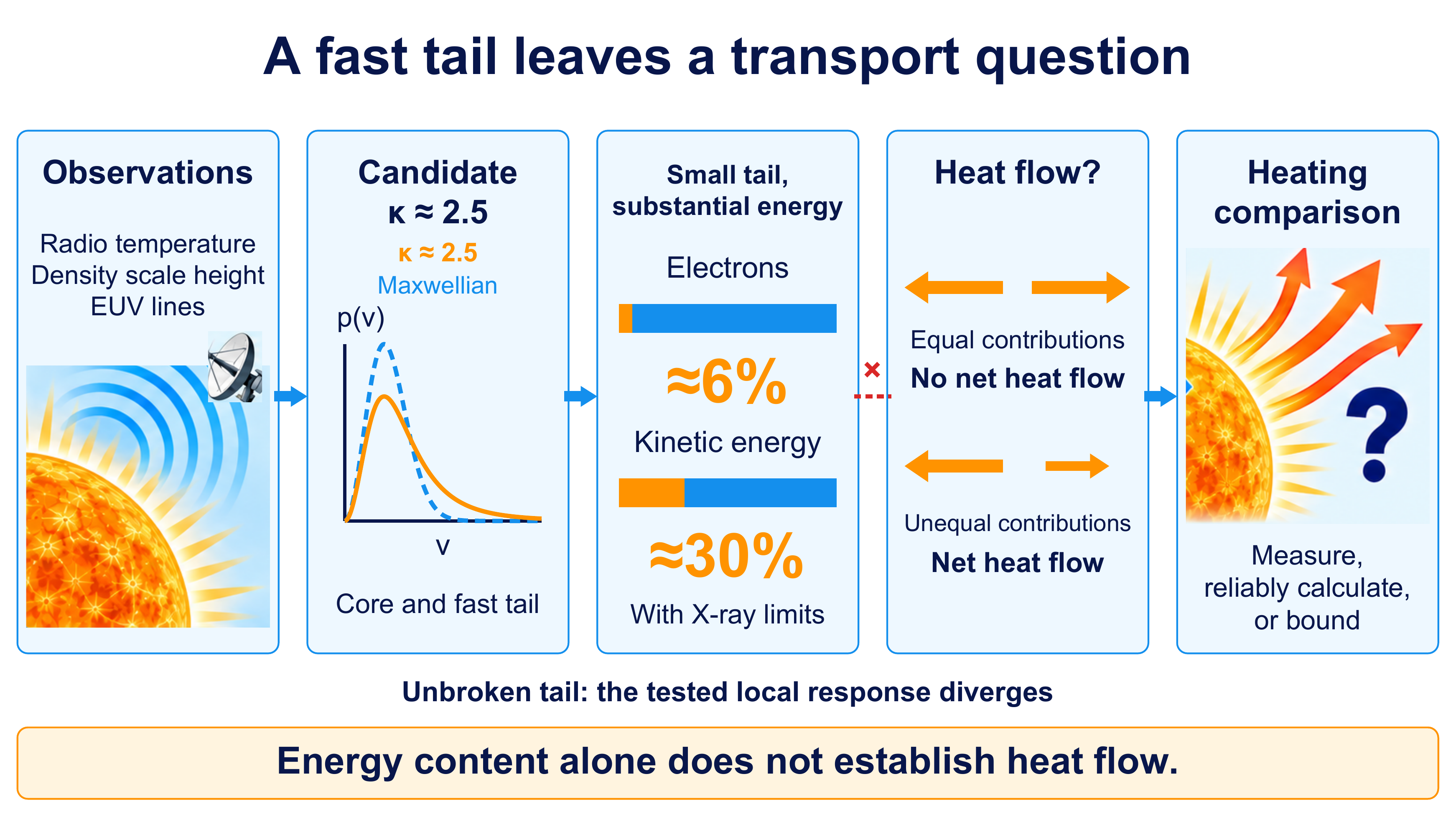}
 \caption{A fast tail can contain substantial energy while leaving
 its heat flow undetermined. In the conditional $\kappa\simeq2.5$
 models restricted by the adopted X-ray limits, about 5--6\%
 of the electrons hold 28--33\% of the kinetic energy. The orange
 segments round these model fractions to 6\% and 30\%; they are
 not measured populations. Opposing arrows show why direction matters:
 equal energy contributions cancel, while unequal contributions leave
 net heat flow. They do not show bulk velocities. For the unbroken
 reference alone, the tested local response requires $\kappa>4$
 (Section~\ref{sec:convergence}). A cutoff removes its divergence
 but does not determine the heat flux. The population must also meet
 the high-energy and maintenance constraints (Section~\ref{sec:origin}).
 Curves, icons, and arrows are schematic. The heating comparison requires
 its heat flow to be established. Decorative illustrations were generated
 with OpenAI image-generation assistance; text and quantitative graphics
 were drawn with a plotting script.}
 \label{fig:kappa-inference}
\end{fixedfigure}

\subsection{Separating a temperature difference from distribution shape}\label{sec:entropy}

The maintenance estimate in Section~\ref{sec:maintenance} needs a measure
of the proposed distribution's departure from a Maxwellian with the
same kinetic energy. Relative entropy supplies that measure. It
separates the difference in shape from the difference between the radio
and mean-energy temperatures. The transport test above depends on a
different, more strongly velocity-weighted integral. For
normalized distributions $f$ and $g$, define
\begin{equation}\label{eq:relative}
 \Dkl(f\Vert g)=\int f\ln(f/g)\,d^3c.
\end{equation}
It is non-negative and vanishes when the distributions agree. With
natural logarithms its dimensionless unit is called a nat. No new
physical unit or statistical confidence level is implied.

Let $M_{\Te}$ be the Maxwellian with the same mean velocity and
kinetic energy as $f$, and let $M_{\Tr}$ have the radio comparison
temperature. Then
\begin{equation}\label{eq:partition}
 \underbrace{\Dkl(f\Vert M_{\Tr})}_{\text{total difference}}
 =\underbrace{\Dkl(f\Vert M_{\Te})}_{\text{shape difference}}
 +\underbrace{\frac32(R-1-\ln R)}_{\text{temperature difference}},
 \qquad R=\frac{\Te}{\Tr}.
\end{equation}
The identity follows because $\ln(M_{\Te}/M_{\Tr})$ is a constant
plus a multiple of $c^2$, whose average is fixed by the matched
energy. Appendix~\ref{app:entropy} gives the derivation and the
$\kappa$ expression. The two terms add because they account for
different parts of one discrepancy. Taking their maximum would
discard one part of that accounting.

The identity does not create a second observation. If the hydrostatic
and radio interpretations justify setting $R=2.4$, the temperature
term is approximately 0.787 nats. Assigning the standard $\kappa$
family then sets $\kappa=2.571$ and supplies a shape term of
approximately 0.293 nats, for a total of 1.080. The representative
$\kappa=2.5$ example has $R=2.5$, giving $1.195=0.320+0.876$ nats.
Those two examples must not be interchanged. Outside the adopted
family, the same ratio does not determine the shape term. If $\Thyd$
is not the electron mean-energy temperature, it cannot be substituted
for $\Te$ in this inference without the required pressure model.

Multiplying the shape term by $\nelec\kb\Te$ gives the
free-energy equivalent
\begin{equation}\label{eq:freeenergy}
 \Delta F=\nelec\kb\Te\Dkl(f\Vert M_{\Te}).
\end{equation}
At fixed density and kinetic energy this is $\Te$ times the entropy
difference from the corresponding Maxwellian. It measures a departure
from equilibrium; it is not extra kinetic energy to be added to the
energy already counted in $f$. Nor is a particular value of relative
entropy a universal threshold for the validity of a transport closure.
The shape difference will be used below to estimate how demanding it
could be to sustain the tail. It does not calculate the conductive
loss to insert into the energy budget. That loss still requires a physical transport
calculation based on equation~(\ref{eq:heatflux}), or an independent
constraint on the flux.

\section{Where could a fast tail come from, and could it be maintained?}\label{sec:origin}

The proposed population must be physically sustainable as well as
able to reproduce temperature readings. The following tests ask
where its fast electrons could come from, how far the tail can extend,
and what is required to maintain it against collisions.

The distribution at a location can reflect both local interactions
and transport from elsewhere. In a weakly collisional atmosphere,
electrons of different energies sample different distances and
encounter different electric and gravitational potentials. Velocity
filtration exploits this selection: a population observed higher in
an atmosphere can have a larger mean energy because slower particles
are less able to reach it, without every surviving particle having
been heated there \citep{scudder1992a,scudder1992b}.

Collisions do not automatically erase the relevance of a boundary
population. In collisional kinetic simulations,
\citet{landi2001} found that sufficiently strong $\kappa$ boundary
tails could support heat flow toward increasing temperature. They
also found approximately classical heat conduction for
$\kappa\gtrsim5$, and required strong boundary tails
($\kappa<4$) to sustain the coronal temperature gradient without
local heating. These are results for their specified boundaries,
not a source solution for the present atmosphere. Other coronal and
transition-region kinetic calculations likewise produce departures
from local Maxwellians \citep{vocks2008,vocks2016}.
In the perturbative wind calculation of \citet{cranmer2014}, tails
were instead bracketed by $\kappa\simeq10$--25; further iteration
was suggested as a possible enhancement, not shown to produce
$\kappa\simeq2.5$.

These results make transport a physical possibility that must meet
source, radiation, and particle-balance constraints. Local production
must supply a driver. The following tests ask whether the proposed
population has those supports. They also separate energy deposited
within the coronal volume from energy carried across its boundary.

\subsection{What does the atmospheric profile constrain?}\label{sec:field}

In the unbroken constant-$\kappa$ collisionless filtration model,
the density and mean-temperature profiles satisfy
\begin{equation}\label{eq:polytrope}
 n\propto T_{\mathrm e}^{-(\kappa-1/2)},\qquad
 \kappa=\frac12-\left(\frac{d\ln T_{\mathrm e}}{d\ln n}\right)^{-1}.
\end{equation}
Applying this relation to the C7 atmospheric
profile calculation \citep{avrett2008} gives $\kappa\simeq2.52$ at
$T\simeq1.08$~MK and a height near 21~Mm. The value ranges
approximately from 2.26 to 3.09 over the selected surrounding interval.
The similarity to the diagnostic example is suggestive within this
model. It is not an independent measurement of $f$: the conversion
of a density--temperature slope to $\kappa$ already assumes the
filtration relation. Other dynamics can generate a similar slope.

The same atmospheric calculation tests a particular local source:
runaway sustained by a weak parallel electric field against Coulomb
drag \citep{dreicer1959,scudder2019serm}. With the adopted
normalization, write the approximate boundary speed as
$x_b=(\alpha/\epsilon)^{1/2}$, where $x_b$ is measured in the
model's thermal-speed convention, $\epsilon$ is the normalized
available field, and $\alpha$ is the drag factor (Appendix~\ref{app:field}).
Near 1~MK, $\epsilon\simeq0.0057$. Taking $\alpha=2$ requires
$\epsilon\simeq0.32$ or 0.22 to reach $x_b=2.5$ or 3,
respectively: roughly 56 or 39 times the available value. The profile
therefore does not supply the field required by this particular
local-runaway construction at those speeds. It does not exclude
waves, intermittent acceleration, or other local energy inputs.

The atmospheric resemblance and the field test answer different
questions. The former gives a conditional shape comparison; the
latter leaves this local production channel short of the required
field. A different local driver or a transported population would
need its own calculation. The field shortfall does not select between
them.

\subsection{Hard X-rays restrict the high-energy continuation}\label{sec:xray}

A power-law fit at lower energies cannot be continued indefinitely
without checking what it would radiate. Quiet-Sun hard-X-ray upper
limits provide a stringent test \citep{hannah2010,foxsi3}. We apply
the adopted Kramers free-free model with emission-measure normalization
$7.6\times10^{47}$~cm$^{-3}$ to $\kappa=2.5$ energy distributions,
using flat averages across the reported photon bands. Appendix
\ref{app:photon} specifies the integrals and the limitations of this
comparison.

The unbroken distribution with $\Tc=0.62$~MK exceeds the adopted
3--6, 6--12, and 5--10~keV band ceilings by factors of approximately
$2.0\times10^3$, $6.7\times10^3$, and $2.0\times10^4$,
respectively. Within this normalization and emission model, it must
steepen substantially. To illustrate the restriction, we examine a
sharp cutoff and distributions continued above a break energy $E_b$
by an exponential in the electron number distribution per unit energy,
with e-folding width
$w$. The core parameter is adjusted so that the terminated radio
source temperature remains $\Tr=0.62$~MK. We require the model
mean-to-radio temperature ratio to lie in 2.1--2.7, the adopted
observational comparison interval.

\begin{table}
\centering
\caption{Conditional high-break intervals for $\kappa=2.5$ at fixed
radio source temperature 0.62~MK. The bounds combine the stated
temperature-ratio interval and flat photon-band ceilings; they are
not empirical confidence intervals.}
\label{tab:edge}
\begin{tabular}{lcc}
\toprule
Continuation & Width $w$ (keV) & Allowed $E_b$ (keV)\\
\midrule
Sharp cutoff & --- & 1.741--3.036\\
Exponential & 0.13 & 1.588--2.515\\
Exponential & 0.20 & 1.477--1.946\\
\bottomrule
\end{tabular}
\end{table}

Table~\ref{tab:edge} preserves a few-keV restriction for these
families while showing that its details depend on the continuation.
In particular, the 0.20~keV example rules out treating 0.15~keV as
a universal maximum edge width. Varying the photon normalization by
a factor of three in either direction moves the sharp-cutoff upper
bound through 3.020--3.063~keV and the $w=0.13$~keV upper bound
through 2.347--2.682~keV. This scalar sensitivity does not replace
folding the spectra through a detector response. Nor do observations
from different epochs and sampled regions constitute a simultaneous
measurement of one electron population.

The location of a restriction does not establish its cause. For
example, cold-target Coulomb slowing over a column \citep{emslie1978} gives
$E_i=(E^2+E_{\mathrm{stop}}^2)^{1/2}$, where $E_i$ is the injected
energy and $E_{\mathrm{stop}}$ the stopping scale. Conservation of
spectral number flux gives
\begin{equation}\label{eq:coldtarget}
 F_N(E)=F_0(E_i)\frac{E}{E_i}.
\end{equation}
For the tested $\kappa$ source and $E_{\mathrm{stop}}=2.5$~keV,
the transmitted-to-unattenuated ratio is about 0.41 at 3~keV,
0.90 at 10~keV, and 0.97 at 20~keV. This process preferentially
removes slower electrons and retains the high-energy continuation.
It does not generate the required upper cutoff. A transport scale
near the inferred break is therefore insufficient evidence for a
common physical origin. Acceleration, escape, and more complete
transport remain candidate ingredients, not calculated explanations.

\subsection{A collisional maintenance estimate}\label{sec:maintenance}

Collisions tend to erase a non-Maxwellian shape. A
single-timescale estimate compares the free-energy equivalent in
equation~(\ref{eq:freeenergy}) with a representative collisional time.
For the sharp-cutoff endpoints in table~\ref{tab:edge}, the mean
temperature is 1.302--1.397~MK and the shape relative entropy is
0.1466--0.2008 nats. At $\nelec=10^8$~cm$^{-3}$ this gives
$\Delta F=0.00263$--$0.00387$~erg~cm$^{-3}$. These examples occupy the lower part of the temperature-ratio
comparison interval; they do not constitute a new fit to the complete
spectrum. The values use
the terminated distributions consistently, including their slightly
adjusted core parameters, 0.62084 and 0.62023~MK.

We evaluate $L\Delta F/\tau$ for an illustrative uniform column of
length $L=50$~Mm. The speed ``three thermal speeds'' needs a
temperature convention. We report both $3\sqrt{2\kb\Tc/\me}$
and $3\sqrt{2\kb\Te/\me}$, and two distinct times: the cold-target
energy-loss time $E/|\dot E|$ and the adopted thermal-mean-free-path
time extrapolated as $v^3$. The latter is defined explicitly in
Appendix~\ref{app:power}; it is not the same collision-time definition
as the former. Table~\ref{tab:power} gives the resulting estimates.

\begin{table}
\centering
\caption{Column-integrated maintenance proxies $L\Delta F/\tau$
in erg~cm$^{-2}$~s$^{-1}$, for $\nelec=10^8$~cm$^{-3}$ and
$L=50$~Mm. Ranges span the two sharp-cutoff endpoints. The columns
are different timescale approximations, not statistical error bounds.}
\label{tab:power}
\begin{tabular}{lcc}
\toprule
Speed & Cold-target loss time & Mean-free-path extrapolation\\
\midrule
$3\sqrt{2\kb\Tc/\me}$ & $(1.81$--$2.66)\times10^7$ & $(2.41$--$3.55)\times10^6$\\
$3\sqrt{2\kb\Te/\me}$ & $(5.95$--$7.88)\times10^6$ & $(0.793$--$1.049)\times10^6$\\
\bottomrule
\end{tabular}
\end{table}

For comparison, applying equation~(\ref{eq:SH}) over the same length
with $|\partial T/\partial s|=T/L$ gives approximately
$3.68\times10^4$~erg~cm$^{-2}$~s$^{-1}$ at 0.62~MK and
$8.10\times10^5$ at 1.5~MK. An assumed radiative coefficient
$10^{-22}$~erg~cm$^3$~s$^{-1}$ and $n_{\mathrm H}=0.83\nelec$
give a column loss of $4.15\times10^3$~erg~cm$^{-2}$~s$^{-1}$.
These are matched illustrative benchmarks; the radiative coefficient
has not been recalculated for the proposed distribution.

These proxies are comparable to or larger than the illustrative
conductive benchmarks. Sustaining the proposed tail is therefore
a substantial physical obligation, and an inexpensive replacement
for the conventional energy supply has not been demonstrated.
The interpretation of the proxy nevertheless requires care: electron self-collisions conserve total electron kinetic
energy while relaxing distribution shape. Dividing its free-energy
equivalent by a test-particle time does not calculate the actual power
of a driver maintaining the full distribution. That requires the
collision operator, driving, and boundary transport together. The
estimate identifies a quantitative obligation for a proposed source;
it neither supplies that source nor revises the observed heating
requirement by the tabulated amount. A maintenance calculation that
rules out this population would remove it as a viable atmospheric
model; that outcome must be allowed in assessing the interpretation.

\subsection{How much source material would be needed?}\label{sec:source}

A transported tail is constrained by what its source would emit.
\citet{anderson1996} provide an important example. Their collisionless
filtration model A, with electron and proton $\kappa=2.5$, boundary
mean temperature 14\,000~K and electron density
$3\times10^{10}$~cm$^{-3}$, overpredicted the tabulated
O\,\textsc{v} 554~\AA\ and O\,\textsc{vi} 1034~\AA\ radiances
by factors of approximately 92 and 6.7. It simultaneously
underpredicted Mg\,\textsc{x} 614~\AA\ by a factor of about 65.
Multiplying the emission to match Mg\,\textsc{x} would increase
the O\,\textsc{v} excess to approximately 6000. A normalization
change cannot reconcile those lines in that model.

This is a physical objection that a new filtration model must meet.
It is not a numerical exclusion of every possible $10^5$~K source:
the boundary distribution, temperature convention, geometry, and
transport must first be specified. For example, applying the
unbroken constant-$\kappa$ relation in
equation~(\ref{eq:polytrope}) to a corona with
$\nelec=10^8$~cm$^{-3}$ and $\Te=1.5$~MK gives a source density
$2.25\times10^{10}$~cm$^{-3}$ if $10^5$~K denotes the source
mean-energy temperature. If it denotes the $\kappa=2.5$ core
parameter instead, the source mean temperature is
$2.5\times10^5$~K and the density is $3.6\times10^9$~cm$^{-3}$.
These are conditional scalings of an unbroken collisionless model,
not density determinations for the terminated examples or synthetic
UV radiances.

The illustrative coronal column used above contains
$N_{\mathrm{tail}}=2.56$--$2.84\times10^{16}$ tail electrons
cm$^{-2}$ above the specified threshold. Converting that inventory
into an amount of source material requires a supply model. For steady
actual particle replacement, number balance would give
\begin{equation}\label{eq:inventory}
 N_{\mathrm s}=N_{\mathrm{tail}}
 \frac{A_{\mathrm c}}{A_{\mathrm s}}
 \frac{\tau_{\mathrm s}}{\eta\tau_{\mathrm{tail}}},
\end{equation}
where $N_{\mathrm s}$ is source electron column, $A_{\mathrm c}$
and $A_{\mathrm s}$ are connected coronal and source areas,
$\eta$ is the fraction of processed source electrons delivered into
the selected tail, and the two times specify source processing and
tail replacement. Shape relaxation does not itself establish a
particle replacement rate. The observations used here do not fix
these quantities, so they supply neither a unique source column nor
an O\,\textsc{v}/O\,\textsc{vi} prediction. The historical UV
conflict and the maintenance estimates consequently remain
constraints on the candidate, not completed checks in its favour.
Fitting the temperature readings has not established the energy and
particle supply needed to sustain the tail. These tests can reject a
proposed tail or identify what its source must supply. They do not
determine the heat carried away by electrons. That requires a flux
measurement, a reliable transport calculation, or a useful bound for
the atmosphere that survives the tests.

\section{What is needed to establish the energy requirement?}\label{sec:conclusion}

For the conductive estimate examined here, the answer to the title
question is no. The temperature observations do not establish whether
the calculated electron heat flow is accurate enough to judge a heating
mechanism's energy supply.

The $\kappa$ family can give one population a cool radio reading and
a higher pressure temperature. The Fe\,\textsc{ix} and radio tests
do not select it uniquely: calibration, temperature structure, and
delayed ion formation leave Maxwellian explanations open. Neither
interpretation reproduces the joint radio, spatial, and spectroscopic
constraints here.

Even a successful reconstruction leaves the heat-flow question.
The tested unbroken-$\kappa$ local response diverges despite finite
pressure. Applying the adopted X-ray limits restricts the tail while
leaving about 5--6\% of the electrons holding 28--33\% of the kinetic
energy. Their heat flow still depends on direction and response to
the surroundings. Source and maintenance tests ask whether this
population can exist; they could rule it out. Calculating its heat
flow is a further physical task.

The observational test should combine radio imaging and spectroscopy
of the same region at the same time. One density distribution and
geometry must account for beam filling, propagation, calibration, and
ion pressure. Absolute line intensities must be tested alongside
ratios, including time-dependent ion formation. A proposed kinetic
atmosphere must also meet the UV and hard-X-ray constraints on its
source and tail. These tests would give a stronger basis for
calculating its heat flow.

Independent constraints on heat flow could settle the budget without
uniquely reconstructing the electrons. Transition-region radiation
can help if the other terms in its energy balance are sufficiently
constrained. Solar-wind distributions offer a separate test: compare
their heat flux before and after reducing them to diagnostic quantities.
Applying that test to the quiet corona would still require the relevant
collision rates and geometry. The viewing dependence of
transition-region Si\,\textsc{iv} profiles reported by
\citet{testa2026} also shows why direction matters, although those
ion measurements do not determine the coronal electron heat flux.

Excluding $\kappa=2.5$ would remove that candidate;
the successful alternative would still need its own justified
transport. Conclusions that remain sound across reliable flux bounds
would survive.

We cannot judge whether a heating mechanism supplies enough energy against a
heat loss we have not established. Until the relevant heat flow is
measured, reliably calculated, or bounded well enough to decide the
comparison, its contribution to the quiet-Sun heating requirement
remains unestablished by the temperature observations examined here.

\section*{Acknowledgements}

The author thanks researchers at
NASA Goddard Space Flight Center for discussions of electron
distributions, diagnostic inference, and transport; E. Landi for
correspondence about atmospheric model data; and the KAPPA, CHIANTI,
instrument, and archive teams for the resources used in this work.

Claude (Anthropic) and Codex (OpenAI) tools assisted with manuscript
organization, analysis and computational checks, figures, and LaTeX formatting.

\section*{Data availability}

No new observations were acquired. The analysis code and reproduction
instructions are available in the accompanying
Zenodo archive (\href{https://doi.org/10.5281/zenodo.21705439}{doi:10.5281/zenodo.21705439}).
External datasets are obtained separately from the sources identified
in the archive. Download scripts and preparation instructions specify
the inputs needed to reproduce the manuscript's numerical results
and tables.

\section*{Funding}

This research received no funding.

\section*{Declaration of interests}

The author reports no conflict of interest.

\appendix

\section{Relative entropy and the temperature partition}\label{app:entropy}

For the zero-mean Maxwellian
$M_T=(\me/2\pi\kb T)^{3/2}\exp[-\me c^2/(2\kb T)]$,
write
\begin{equation}
 \Dkl(f\Vert M_{\Tr})-\Dkl(f\Vert M_{\Te})
 =\int f\ln\frac{M_{\Te}}{M_{\Tr}}\,d^3c.
\end{equation}
Normalization and $\langle\me c^2/2\rangle_f=3\kb\Te/2$ reduce
the right-hand side to
$\tfrac32[\Te/\Tr-1-\ln(\Te/\Tr)]$, establishing
equation~(\ref{eq:partition}). This is the moment-matched
exponential-family identity in a directly evaluable form
\citep{csiszar1975,covthomas2006}. It holds for any normalized $f$
with the required finite integrals, not only a $\kappa$ distribution.
It also shows why matching energy removes the temperature term from
the shape comparison.

For the unbroken distribution of equation~(\ref{eq:kappa}), direct
integration gives
\begin{align}\label{eq:analyticshape}
 \Dkl(f_\kappa\Vert M_{\Te})={}&
 \ln\frac{\Gamma(\kappa+1)}{\Gamma(\kappa-1/2)}
 -\frac32\ln(\kappa-3/2)+\frac32\nonumber\\
 &-(\kappa+1)\big[\psi(\kappa+1)-\psi(\kappa-1/2)\big],
\end{align}
where $\psi(z)=d\ln\Gamma(z)/dz$. The finite-energy condition is
$\kappa>3/2$. The corresponding identity between entropy deficit
and free-energy difference is standard at the stated reference
temperature \citep{esposito2011,parrondo2015}; its use here does
not prescribe a physical reservoir or a maintenance mechanism.

For the sharp-cutoff models, the distribution is renormalized below
the cutoff, its mean energy is recomputed, and the two relative
entropies are integrated directly. At the fixed-radio endpoints of
table~\ref{tab:edge}, the partition relative to $M_{0.62\,\mathrm{MK}}$
is, in nats,
\begin{equation}
 0.68365=0.14656+0.53709,\qquad
 0.86195=0.20085+0.66111.
\end{equation}
The shape free-energy equivalents are 9.77\% and 13.39\% of the
kinetic energy. They are different quantities from the fractions of
energy contained above a speed threshold. In computing those latter
fractions, the threshold is kept at 2.1 times the thermal speed of
the corresponding unbroken mean temperature, or 3.3204 times the
core speed $\sqrt{2\kb\Tc/\me}$. It is not moved to the lower
mean temperature of the terminated distribution.

\section{Spectroscopic test definitions and retained checks}\label{app:feix}

The primary test uses $y=I(197.862)/I(177.592)$ in energy units,
with the calibration check $x=I(189.941)/I(177.592)$ in the same
units. The separate $r_{197}$ comparison in the main text uses
photon units; wavelength conversion is necessary before identifying
it with $y/x$. The observed-wavelength labels differ slightly from
the atomic table values, 197.854 and 189.935~\AA. The latter identify
the same transitions in the 915-level model
\citep{dzsbm2014,dz23}.

The baseline calibration-check prediction is $x=1.42\pm0.10$.
Measured values are $1.771\pm0.339$ in the atlas region and
$2.111\pm0.223$ in the inner annulus. The clear inner-annulus
failure prevents a distribution claim from the primary test. Under
the 2013 calibration, that check passes but the inner-annulus primary
residual is approximately 2.6 adopted systematic-allowance units,
below the specified threshold of 3. The requirement that the
calibration check pass is retained in interpreting every primary
comparison.

The nearby-line baseline measurements are
$r_{197}=1.291\pm0.018$ and $1.362\pm0.005$, where these errors
are statistical only. Density, calibration, atomic data, and
temperature structure introduce additional uncertainty. The
9\% combined allowance and the alternative empirical
allowances are retained separately; reducing an allowance does not
remove a modelling degeneracy. The broad emission-measure example
uses the distribution recovered from \citet{dzww2025}, with a
Maxwellian reference ratio 1.13178 at
$\nelec=2\times10^8$~cm$^{-3}$. The narrow reference is
0.988885 at $\log_{10}(\nelec/\mathrm{cm}^{-3})=8.4$ and peak
$\log_{10}(T/\mathrm K)=6.05$. Neither is a universal Maxwellian
prediction. Applying a Maxwellian-derived emission-measure
distribution to non-Maxwellian emissivities is an additional
approximation; a fully consistent comparison would refit it.

The raster identifier is \texttt{eis\_l0\_20070311\_023212}.
The reduction pools drift-corrected spectra and fits
multi-Gaussian line groups with a linear background, explicitly
including neighbouring features, using EISPAC
\citep{weberg2023}. The outer annulus was not scored because of
contamination. A separate 2010 raster was not scored because the
working deblends failed quality checks, with reduced residual
statistics of order $10^3$. These are disclosed data-quality
decisions; they are not described as prospective selection rules.
The atomic-data sensitivity of the 197.862~\AA\ transition and the
short-wavelength calibration remain in the retained execution
record. Neither is removed by the same-ion construction.

Additional population comparisons are contextual. Across 25 on-disk
regions, the mean ratio was $1.180\pm0.088$; the Pearson
correlation with the adopted density estimate was $-0.68$, while
the rank correlation was $-0.27$. Removing the three densest points
reduced the Pearson value to approximately 0.02. This is not robust
evidence of a population law. Across 87 pointings, a raw detector
ratio near 1.478 with about 8\% scatter was stable across the
examined epochs. Stability of an uncalibrated count ratio is useful
instrumental context, but is not an absolute electron-distribution
measurement.

For the new excitation-envelope test, 377\,628 collision transitions
are included. The level-population balance residual is below
$7\times10^{-15}$, and the narrow and broad reference ratios above
are reproduced. At fixed statistical-equilibrium emissivities, an
integrated line ratio is a positive, denominator-emissivity-weighted
average of local ratios. The excitation envelope thus gives an
attainability test when the Fe\,\textsc{ix} abundance is left free.
It does not determine the temporal ion abundance, absolute emission
measure, or duration of a proposed event.

\subsection{Density conditioning and the treatment comparison}\label{app:feix-treatments}

The Fe\,\textsc{xiii} (203.797+203.828)/202.044~\AA\ ratio
provides density estimates from the same raster. The atlas-region
$\log_{10}(\nelec/\mathrm{cm}^{-3})$ values are 8.51, 8.44,
8.42, and 8.36 under Maxwellian, $\kappa=3$, 2.5, and 2
assumptions, respectively; the corresponding inner-annulus estimates
span 8.19--8.34. An Fe\,\textsc{xii} ratio gives values within
0.08 dex in the compared reductions. This agreement supports the
density conditioning of the comparison, while retaining the
assumptions about how the different ions sample the atmosphere.

Table~\ref{tab:feix-treatments} preserves the main treatment
dependence. Each entry is an observed-minus-model residual divided
by a specified allowance, including the model variation used in
the original comparison. A magnitude below three meets the adopted
consistency rule. The empirical allowances of 3\% and 4\% for the
nearby-line pair summarize extraction variants in the atlas and
annulus, respectively. They do not independently constrain every
calibration, atomic-data, or temperature-structure error. The 9\%
allowance is therefore retained as a separate treatment.

\begin{table}
\centering
\caption{Retained Fe\,\textsc{ix} comparison. Paired entries are
atlas/inner-annulus residuals in the stated allowance units.
The first four rows use narrow temperature conditioning and
each hypothesis's density; the last two use the broad emission
measure distribution at $2\times10^8$~cm$^{-3}$ in the atlas region.
These are sensitivity comparisons, not Gaussian significances.}
\label{tab:feix-treatments}
\small
\begin{tabular}{lrrrr}
\toprule
Treatment & Maxwellian & $\kappa=3$ & $\kappa=2.5$ & $\kappa=2$\\
\midrule
Baseline $r_{197}$, 9\% & $2.8/2.8$ & $0.5/0.5$ & $-0.6/-0.7$ & $-3.4/-3.4$\\
Baseline $r_{197}$, empirical & $6.6/5.7$ & $1.2/1.0$ & $-1.4/-1.3$ & $-5.1/-5.0$\\
2013 $y$, 9\% & $0.8/2.6$ & $-0.5/0.7$ & $-1.1/-0.3$ & $-2.8/-2.7$\\
2013 $r_{197}$, empirical & $5.0/4.3$ & $-0.9/-0.7$ & $-3.2/-2.9$ & $-6.2/-6.0$\\
Broad DEM $r_{197}$, 9\% & 1.4 & 0.4 & $-0.3$ & $-2.4$\\
Broad DEM $r_{197}$, empirical & 3.7 & 1.0 & $-0.7$ & $-6.5$\\
\bottomrule
\end{tabular}
\end{table}

The failed baseline calibration check applies to the primary $y$
test regardless of the nearby-line residuals. In the 2013 treatment,
the live $y$ test remains below the required Maxwellian rejection
threshold. The table therefore retains a conditional preference in
some comparisons without changing the primary test's outcome.

\subsection{Extraction and atomic-channel checks}\label{app:feix-channels}

Independent spatial re-pooling within the two regions gives
$r_{197}=1.288$--1.399. Fits with varied windows, backgrounds, and
width bounds contribute to the empirical allowances above. These
checks share calibration and atomic data, so they test extraction
stability rather than provide independent distribution measurements.

Two channel residuals in table~\ref{tab:feix-channels} are largely
insensitive to the tested distribution choices. The larger off-limb
deficit of 177.592~\AA\ cannot be attributed to one fixed
effective-area correction alone. The 189.572~\AA\ ratio also
remains low in the on-disk comparisons. These identify atomic-data
or extraction questions that persist under the different distribution
hypotheses; they do not assign a unique cause. The clean nearby
188.493~\AA\ line checks the denominator shared by the working ratios.

\begin{table}
\centering
\caption{Selected atlas-region channel checks: observed/model
for each line's intensity relative to 189.935~\AA. Ranges span
Maxwellian and $\kappa$ treatments. They identify residuals to
investigate, rather than independent distribution determinations.}
\label{tab:feix-channels}
\begin{tabular}{lll}
\toprule
Line (\AA) & Observed/model & Role\\
\midrule
177.592 & 0.78--0.79 & Calibration-check deficit\\
188.493 & 1.04--1.11 & Shared-denominator check\\
189.572 & 0.66--0.68 & Persistent channel deficit\\
\bottomrule
\end{tabular}
\end{table}

Doubling the 197.862~\AA\ emissivity to represent the older
oscillator-strength alternative worsens every tested hypothesis in
the scored regions, giving residuals about 4.9--11 times the stated
allowance. This favors the adopted 2014 atomic treatment within
the compared models \citep{dzsbm2014}; it does not eliminate all
remaining atomic uncertainty.

\section{Radio and photon calculations for terminated tails}\label{app:photon}

The unbroken electron energy probability density is
\begin{equation}
 p(E)=C\sqrt E\,(1+E/E_0)^{-\kappa-1},\qquad E_0=\kappa\kb\Tc,
\end{equation}
with $C$ fixed by normalization. A sharp cutoff sets $p(E)=0$
above $E_b$. The exponential family replaces the continuation by
$p(E_b)\exp[-(E-E_b)/w]$, with each complete distribution
renormalized. The exponential is specified for $p(E)$, not for
the phase-space density or injected flux.

The isotropic low-frequency free-free source function used here
follows \citet{fk2014} with slowly varying Gaunt factors. For a
sharp cutoff it gives
\begin{equation}\label{eq:cutradio}
 \kb\Tr=\frac{E_0}{\kappa}
 \left[1-(1+E_b/E_0)^{-\kappa}\right].
\end{equation}
For the exponential continuation, the bracketed integral is replaced
by
\begin{equation}
 \kb\Tr=\int_0^{E_b}(1+E/E_0)^{-\kappa-1}\,dE
 +\sqrt{E_b}(1+E_b/E_0)^{-\kappa-1}
 \int_{E_b}^{\infty}\frac{e^{-(E-E_b)/w}}{\sqrt E}\,dE.
\end{equation}
These expressions tend to $\Tr=\Tc$ for the unbroken case.
For each break, $\Tc$ is solved so that $\Tr=0.62$~MK, and
$\Te=2\langle E\rangle/(3\kb)$ is evaluated from the normalized
terminated $p(E)$. This avoids mixing the legacy energy scale
$E_0=0.1293$~keV with $\Tc=0.62$~MK, which would instead give
$E_0\simeq0.13357$~keV at $\kappa=2.5$.

With photon energy $\varepsilon$, the adopted non-relativistic
Kramers kernel is $d\sigma/d\varepsilon=\sigma_0/(\varepsilon E)$,
where energies are in keV and $\sigma_0=7.9\times10^{-25}$~cm$^2$\,keV.
The photon flux at distance $d=1$~au is
\begin{equation}
 J(\varepsilon)=\frac{\mathrm{EM}}{4\pi d^2}
 \int_{\varepsilon}^{\infty}v(E)p(E)
       \frac{\sigma_0}{\varepsilon E}\,dE.
\end{equation}
Here, EM denotes the adopted density-product volume normalization;
no revised emission-measure inference from the observations is
performed. Averaging across $[a,b]$ replaces $1/\varepsilon$ in
the reversed integral by $\ln[\min(b,E)/a]/(b-a)$ for $E\geq a$.
The adopted ceilings are 0.034 in 3--6~keV and 0.00095 in
6--12~keV from the RHESSI comparison, and 0.0006 in 5--10~keV
from the FOXSI comparison, in photons~cm$^{-2}$~s$^{-1}$~keV$^{-1}$
\citep{hannah2010,foxsi3}. Their observational selections differ;
using them here is a conditional model constraint, not a joint
likelihood for a single observed region.

Numerical integration is split at breaks and band boundaries, with
small dimensional factors applied after quadrature. An independent
change of variable checks the narrow exponential tail, and reversing
the photon/electron integration order checks the band average. These
steps ensure that the narrow high-energy continuation is resolved
before applying the observational ceilings. Instrument-response fitting and
the source geometry would be required for an empirical interval.

\section{Collision-time and atmospheric-field conventions}\label{app:power}

With temperature in kelvin, electron density in cm$^{-3}$, and
$T_{\mathrm{eV}}=\kb T/(1~\mathrm{eV})$, the adopted convention is
\begin{equation}
 \ln\Lambda=24-\ln(\sqrt{\nelec}/T_{\mathrm{eV}}),\qquad
 \lambda_0=\frac{1.07\times10^5T^2}{\nelec\ln\Lambda}\ \mathrm{cm}.
\end{equation}
The maintenance calculation evaluates this at the core temperature
and uses $v_0=\sqrt{\kb\Tc/\me}$, giving
$\tau_\lambda=(\lambda_0/v_0)(v/v_0)^3$. This explicitly retains
the thermal convention used in the atmospheric calculation
\citep{nrlformulary}. For the cold-target comparison,
$dE/dN=-K/E$, $K=2\pi e^4\ln\Lambda$ in Gaussian cgs units, and
$dN/dt=\nelec v$, so
\begin{equation}
 \tau_E=\frac{E}{|\dot E|}
       =\frac{E^2}{2\pi e^4\ln\Lambda\,\nelec v},\qquad
       E=\frac12\me v^2.
\end{equation}
At three core thermal speeds these times are about 5.47 and
0.73~s, respectively. Warm-target diffusion, collective effects,
directional scattering, and the relaxation of the full shape are
not computed by either single-time estimate.

\subsection{Inherited C7 profile calculation}\label{app:field}

The atmospheric-field estimate evaluates
\begin{equation}
 \epsilon=\frac12\frac{\lambda_0}{L_P}
          +0.71\frac{\lambda_0}{L_T},\qquad
 L_P=\left|\frac{d\ln(\nelec\kb T)}{dh}\right|^{-1},\quad
 L_T=\left|\frac{d\ln T}{dh}\right|^{-1}.
\end{equation}
It uses the Dreicer-field normalization $E_D=2\kb T/(\lambda_0|e|)$
and a thermal-force allowance twice the classical 0.355 coefficient
in this convention \citep{scudder2019tf}. That allowance is a model
choice, not a rigorous bound on a kinetic thermal force. The
normalized field ranges from approximately 0.0036 to 0.0117 over
the retained 0.5--1.5~MK column. The 1~MK interpolation is 0.0057;
the 1.08~MK crossing row is 0.0064. Keeping those anchors separate
gives the 39--56 shortfall quoted at 1~MK in the main text.
As a limited prefactor check, tripling $\lambda_0$ triples the
adopted field and reduces those ratios to approximately 13--19.
This sensitivity does not bound an uncalculated kinetic thermal
force or a different acceleration mechanism.

The inherited profile differentiates natural logarithms with
second-order interior differences and one-sided endpoint stencils.
The upper endpoint is excluded from the reported comparison because
its gradient is edge affected. With $x=v/\sqrt{2\kb T/\me}$,
the stated $x_b=(\alpha/\epsilon)^{1/2}$ construction gives
$x_b\simeq17.7$ at the crossing for $\alpha=2$; the published drag
range is approximately 1.42--3. The numerical profile is retained
as a conditional test of this local mechanism. The saved derived
columns permit inspection of the quoted values, while a fresh
reconstruction from the original atmosphere table is outside the
new bounded calculation pass.

\subsection{Numerical implementation}

The new bounded calculations were implemented with NumPy and SciPy
and plotted with Matplotlib. The checks include
independent energy- and velocity-space moments, the analytic entropy
identity, an analytic sharp-cutoff radio expression, and alternative
photon-integration variables and order. A separate check of
equation~(\ref{eq:responseintegral}) uses $u=\sqrt{A}v$:
the dimensionless integral equals $\tfrac12 B(6,\kappa-4)$ for
$\kappa>4$, where $B$ is the beta function. Direct quadrature
agrees at $\kappa=4.5$, 5, and 8 to relative error below
$10^{-9}$, and finite-cutoff evaluations reproduce the power and
logarithmic divergences stated in Section~\ref{sec:convergence}.
This is an integral-domain check, not a new solar heat-flux model.
Reported feasibility tests retain their specified model assumptions;
numerical agreement does not constitute observational validation
of those assumptions.

\begingroup
\small
\bibliographystyle{plainnat}
\bibliography{references}

@ARTICLE{scudder1992a,
  author = {{Scudder}, J.~D.}, year = 1992,
  title = {On the Causes of Temperature Change in Inhomogeneous Low-Density Astrophysical Plasmas},
  journal = {ApJ}, volume = 398, pages = {299}, doi = {10.1086/171858}
}

@ARTICLE{scudder1992b,
  author = {{Scudder}, J.~D.}, year = 1992,
  title = {Why All Stars Should Possess Circumstellar Temperature Inversions},
  journal = {ApJ}, volume = 398, pages = {319}, doi = {10.1086/171859}
}

@ARTICLE{scudder2019tf,
  author = {{Scudder}, J.~D.}, year = 2019,
  title = {The Thermal Force in Astrophysical Plasmas: Current-Free {Coulomb} Friction},
  journal = {ApJ}, volume = 882, pages = {146}, doi = {10.3847/1538-4357/ab3348}
}

@ARTICLE{scudder2019serm,
  author = {{Scudder}, J.~D.}, year = 2019,
  title = {Steady Electron Runaway Model SERM: Astrophysical Alternative for the {Maxwellian} Assumption},
  journal = {ApJ}, volume = 885, pages = {138}, doi = {10.3847/1538-4357/ab4882}
}

@ARTICLE{scudder2021,
  author = {{Scudder}, J.~D.}, year = 2021,
  title = {Quality Metric for {Spitzer}--{Braginskii} and {Grad} 8 Moment Heat Flux Closures},
  journal = {ApJ}, volume = 907, pages = {90}, doi = {10.3847/1538-4357/abc475}
}

@ARTICLE{scudderolbert1979,
  author = {{Scudder}, J.~D. and {Olbert}, S.}, year = 1979,
  title = {A Theory of Local and Global Processes Which Affect Solar Wind Electrons, 1. The Origin of Typical 1 AU Velocity Distribution Functions---Steady State Theory},
  journal = {J. Geophys. Res.}, volume = 84, pages = {2755}, doi = {10.1029/JA084iA06p02755}
}

@ARTICLE{dreicer1959,
  author = {{Dreicer}, H.}, year = 1959,
  title = {Electron and Ion Runaway in a Fully Ionized Gas. I},
  journal = {Phys. Rev.}, volume = 115, pages = {238}, doi = {10.1103/PhysRev.115.238}
}

@ARTICLE{shoub1983,
  author = {{Shoub}, E.~C.}, year = 1983,
  title = {Invalidity of Local Thermodynamic Equilibrium for Electrons in the Solar Transition Region. I. {Fokker}-{Planck} Results},
  journal = {ApJ}, volume = 266, pages = {339}, doi = {10.1086/160783}
}

@ARTICLE{avrett2008,
  author = {{Avrett}, E.~H. and {Loeser}, R.}, year = 2008,
  title = {Models of the Solar Chromosphere and Transition Region from {SUMER} and {HRTS} Observations: Formation of the Extreme-Ultraviolet Spectrum of Hydrogen, Carbon, and Oxygen},
  journal = {ApJS}, volume = 175, pages = {229}, doi = {10.1086/523671}
}

@ARTICLE{landi2001,
  author = {{Landi}, S. and {Pantellini}, F.~G.~E.}, year = 2001,
  title = {On the Temperature Profile and Heat Flux in the Solar Corona: Kinetic Simulations},
  journal = {A\&A}, volume = 372, pages = {686}, doi = {10.1051/0004-6361:20010552}
}

@ARTICLE{vasyliunas1968,
  author = {{Vasyli{\= u}nas}, V.~M.}, year = 1968,
  title = {A Survey of Low-Energy Electrons in the Evening Sector of the Magnetosphere with {OGO} 1 and {OGO} 3},
  journal = {J. Geophys. Res.}, volume = 73, pages = {2839}, doi = {10.1029/JA073i009p02839}
}

@ARTICLE{vocks2008,
  author = {{Vocks}, C. and {Mann}, G. and {Rausche}, G.}, year = 2008,
  title = {Formation of Suprathermal Electron Distributions in the Quiet Solar Corona},
  journal = {A\&A}, volume = 480, pages = {527}, doi = {10.1051/0004-6361:20078826}
}

@ARTICLE{vocks2016,
  author = {{Vocks}, C. and {Dzif{\v c}{\'a}kov{\'a}}, E. and {Mann}, G.}, year = 2016,
  title = {Suprathermal Electron Distributions in the Solar Transition Region},
  journal = {A\&A}, volume = 596, pages = {A41}, doi = {10.1051/0004-6361/201629209}
}

@ARTICLE{hannah2010,
  author = {{Hannah}, I.~G. and {Hudson}, H.~S. and {Hurford}, G.~J. and {Lin}, R.~P.}, year = 2010,
  title = {Constraining the Hard X-Ray Properties of the Quiet {Sun} with New {RHESSI} Observations},
  journal = {ApJ}, volume = 724, pages = {487}, doi = {10.1088/0004-637X/724/1/487}
}

@ARTICLE{foxsi3,
  author = {{Buitrago-Casas}, J.~C. and {Glesener}, L. and {Christe}, S. and
            {Krucker}, S. and {Vievering}, J. and {Athiray}, P.~S. and others}, year = 2022,
  title = {On the Faintest Solar Coronal Hard X-Rays Observed with {FOXSI}},
  journal = {A\&A}, volume = 665, pages = {A103}, doi = {10.1051/0004-6361/202243272}
}

@ARTICLE{sharma2020,
  author = {{Sharma}, R. and {Oberoi}, D.}, year = 2020,
  title = {Propagation Effects in Quiet {Sun} Observations at Meter Wavelengths},
  journal = {ApJ}, volume = 903, pages = {126}, doi = {10.3847/1538-4357/abb949}
}

@ARTICLE{cranmer2014,
  author = {{Cranmer}, S.~R.}, year = 2014,
  title = {Suprathermal Electrons in the Solar Corona: Can Nonlocal Transport Explain Heliospheric Charge States?},
  journal = {ApJ}, volume = 791, pages = {L31}, doi = {10.1088/2041-8205/791/2/L31}
}

@ARTICLE{dzifcakova2013,
  author = {{Dzif{\v c}{\'a}kov{\'a}}, E. and {Dud{\'i}k}, J.}, year = 2013,
  title = {H to Zn Ionization Equilibrium for the Non-{Maxwellian} Electron $\kappa$-distributions: Updated Calculations},
  journal = {ApJS}, volume = 206, pages = {6}, doi = {10.1088/0067-0049/206/1/6}
}

@ARTICLE{mercier2015,
  author = {{Mercier}, C. and {Chambe}, G.}, year = 2015,
  title = {Electron Density and Temperature in the Solar Corona from Multifrequency Radio Imaging},
  journal = {A\&A}, volume = 583, pages = {A101}, doi = {10.1051/0004-6361/201425540}
}

@ARTICLE{lorincik2020,
  author = {{L{\"o}rin{\v c}{\'i}k}, J. and {Dud{\'i}k}, J. and {Del Zanna}, G. and
            {Dzif{\v c}{\'a}kov{\'a}}, E. and {Mason}, H.~E.}, year = 2020,
  title = {Plasma Diagnostics from Active Region and Quiet-{Sun} Spectra Observed by {Hinode}/{EIS}: Quantifying the Departures from a {Maxwellian} Distribution},
  journal = {ApJ}, volume = 893, pages = {34}, doi = {10.3847/1538-4357/ab8010}
}

@ARTICLE{edmonds2026a,
  author = {{Edmonds}, V.}, year = 2026,
  title = {The Diagnostic Temperature Discrepancy as Evidence for Non-{Maxwellian} Coronal Electrons},
  journal = {Open J. Astrophys.}, volume = 9, doi = {10.33232/001c.161223}
}

@ARTICLE{mercier2016corr,
  author = {{Mercier}, C. and {Chambe}, G.}, year = 2016,
  title = {Electron Density and Temperature in the Solar Corona from Multifrequency Radio Imaging ({Corrigendum})},
  journal = {A\&A}, volume = 585, pages = {C1},
  doi = {10.1051/0004-6361/201425540e}
}

@ARTICLE{edmonds2026conv,
  author = {{Edmonds}, V.}, year = 2026,
  title = {Multi-diagnostic Convergence: A Single Measurement in Weakly Collisional Plasmas},
  journal = {Open Transport}, volume = 1, number = 1, pages = {20260011}, doi = {10.1515/ot-2026-0011}
}

@ARTICLE{edmonds2026dem,
  author = {{Edmonds}, V.}, year = 2026,
  title = {The Quiet-{Sun} {DEM} Under Kappa: Diagnostic Degeneracy and the Failure of the Conductive Closure},
  journal = {Transport Phenomena}, volume = 1, number = 3, pages = {20260065},
  doi = {10.1515/tp-2026-0065}
}

@ARTICLE{braginskii1965,
  author = {{Braginskii}, S.~I.}, year = 1965,
  title = {Transport Processes in a Plasma},
  journal = {Rev. Plasma Phys.}, volume = 1, pages = {205}
}

@BOOK{covthomas2006,
  author = {{Cover}, T.~M. and {Thomas}, J.~A.}, year = 2006,
  title = {Elements of Information Theory}, edition = {2nd},
  publisher = {Wiley}, address = {Hoboken, NJ}
}

@ARTICLE{cranmerschiff2021,
  author = {{Cranmer}, S.~R. and {Schiff}, A.~J.}, year = 2021,
  title = {Electron Heat Flux in the Solar Wind: Generalized Approaches to Fluid Transport with a Variety of Skewed Velocity Distributions},
  journal = {J. Geophys. Res. Space Physics}, volume = 126, pages = {e2021JA029666},
  doi = {10.1029/2021JA029666}
}

@ARTICLE{csiszar1975,
  author = {{Csisz{\'a}r}, I.}, year = 1975,
  title = {$I$-Divergence Geometry of Probability Distributions and Minimization Problems},
  journal = {Ann. Probab.}, volume = 3, pages = {146}, doi = {10.1214/aop/1176996454}
}

@ARTICLE{du2013,
  author = {{Du}, J.}, year = 2013,
  title = {Transport Coefficients in {Lorentz} Plasmas with the Power-law Kappa-distribution},
  journal = {Phys. Plasmas}, volume = 20, pages = {092901}, doi = {10.1063/1.4820799}
}

@ARTICLE{dudik2014,
  author = {{Dud{\'i}k}, J. and {Del Zanna}, G. and {Mason}, H.~E. and {Dzif{\v c}{\'a}kov{\'a}}, E.}, year = 2014,
  title = {Signatures of the Non-{Maxwellian} $\kappa$-distributions in Optically Thin Line Spectra. I. Theory and Synthetic {Fe IX--XIII} Spectra},
  journal = {A\&A}, volume = 570, pages = {A124}, doi = {10.1051/0004-6361/201424124}
}

@ARTICLE{dz23,
  author = {{Dzif{\v c}{\'a}kov{\'a}}, E. and {Dud{\'i}k}, J. and {Pavelkov{\'a}}, M. and
            {Solarov{\'a}}, B. and {Zemanov{\'a}}, A.}, year = 2023,
  title = {{KAPPA}: A Package for the Synthesis of Optically Thin Spectra for the Non-{Maxwellian} $\kappa$-distributions. III. Improvements to Ionization Equilibrium and Extension to $\kappa < 2$},
  journal = {ApJS}, volume = 269, pages = {45}, doi = {10.3847/1538-4365/ad014d}
}

@ARTICLE{esposito2011,
  author = {{Esposito}, M. and {Van den Broeck}, C.}, year = 2011,
  title = {Second Law and {Landauer} Principle Far from Equilibrium},
  journal = {EPL}, volume = 95, pages = {40004}, doi = {10.1209/0295-5075/95/40004}
}

@ARTICLE{fk2014,
  author = {{Fleishman}, G.~D. and {Kuznetsov}, A.~A.}, year = 2014,
  title = {Theory of Gyroresonance and Free-Free Emissions from Non-{Maxwellian} Quasi-steady-state Electron Distributions},
  journal = {ApJ}, volume = 781, pages = {77}, doi = {10.1088/0004-637X/781/2/77}
}

@ARTICLE{husidic2021,
  author = {{Husidic}, E. and {Lazar}, M. and {Fichtner}, H. and {Scherer}, K. and {Poedts}, S.}, year = 2021,
  title = {Transport Coefficients Enhanced by Suprathermal Particles in Nonequilibrium Heliospheric Plasmas},
  journal = {A\&A}, volume = 654, pages = {A99}, doi = {10.1051/0004-6361/202141760}
}

@ARTICLE{husidic2022,
  author = {{Husidic}, E. and {Scherer}, K. and {Lazar}, M. and {Fichtner}, H. and {Poedts}, S.}, year = 2022,
  title = {Toward a Realistic Evaluation of Transport Coefficients in Non-equilibrium Space Plasmas},
  journal = {ApJ}, volume = 927, pages = {159}, doi = {10.3847/1538-4357/ac4af4}
}

@ARTICLE{levermore1996,
  author = {{Levermore}, C.~D.}, year = 1996,
  title = {Moment Closure Hierarchies for Kinetic Theories},
  journal = {J. Stat. Phys.}, volume = 83, pages = {1021}, doi = {10.1007/BF02179552}
}

@ARTICLE{owockiscudder1983,
  author = {{Owocki}, S.~P. and {Scudder}, J.~D.}, year = 1983,
  title = {The Effect of a Non-{Maxwellian} Electron Distribution on Oxygen and Iron Ionization Balances in the Solar Corona},
  journal = {ApJ}, volume = 270, pages = {758}, doi = {10.1086/161167}
}

@ARTICLE{parrondo2015,
  author = {{Parrondo}, J.~M.~R. and {Horowitz}, J.~M. and {Sagawa}, T.}, year = 2015,
  title = {Thermodynamics of Information},
  journal = {Nat. Phys.}, volume = 11, pages = {131}, doi = {10.1038/nphys3230}
}

@ARTICLE{spitzer1953,
  author = {{Spitzer}, L. and {H{\"a}rm}, R.}, year = 1953,
  title = {Transport Phenomena in a Completely Ionized Gas},
  journal = {Phys. Rev.}, volume = 89, pages = {977}, doi = {10.1103/PhysRev.89.977}
}

@ARTICLE{withbroe1977,
  author = {{Withbroe}, G.~L. and {Noyes}, R.~W.}, year = 1977,
  title = {Mass and Energy Flow in the Solar Chromosphere and Corona},
  journal = {ARA\&A}, volume = 15, pages = {363}, doi = {10.1146/annurev.aa.15.090177.002051}
}

@ARTICLE{emslie1978,
  author = {{Emslie}, A.~G.}, year = 1978,
  title = {The Collisional Interaction of a Beam of Charged Particles with a Hydrogen Target of Arbitrary Ionization Level},
  journal = {ApJ}, volume = 224, pages = {241}, doi = {10.1086/156371}
}

@MISC{nrlformulary,
  author = {{Beresnyak}, A.}, year = 2023,
  title = {{NRL} Plasma Formulary},
  howpublished = {Naval Research Laboratory, Washington, DC}
}

@ARTICLE{delzanna2012,
  author = {{Del Zanna}, G.}, year = 2012,
  title = {Benchmarking Atomic Data for the {CHIANTI} Atomic Database: Coronal Lines Observed by {Hinode} {EIS}},
  journal = {A\&A}, volume = 537, pages = {A38}, doi = {10.1051/0004-6361/201117592}
}

@ARTICLE{dzsbm2014,
  author = {{Del Zanna}, G. and {Storey}, P.~J. and {Badnell}, N.~R. and {Mason}, H.~E.}, year = 2014,
  title = {Atomic Data for Astrophysics: {Fe IX}},
  journal = {A\&A}, volume = 565, pages = {A77}, doi = {10.1051/0004-6361/201323297}
}

@ARTICLE{cranmergilly2026,
  author = {{Cranmer}, S.~R. and {Gilly}, C.~R.},
  year = 2026,
  title = {Testing Theories of Solar Coronal Heating with Three-Dimensional Forward Modeling},
  journal = {ApJ},
  volume = {1008},
  pages = {49},
  doi = {10.3847/1538-4357/ae8f30}
}

@ARTICLE{delzanna2010,
  author = {{Del Zanna}, G.}, year = 2010,
  title = {Benchmarking Atomic Data for Astrophysics: {Fe XI}},
  journal = {A\&A}, volume = 514, pages = {A41}, doi = {10.1051/0004-6361/201014063}
}

@ARTICLE{delzanna2013cal,
  author = {{Del Zanna}, G.}, year = 2013,
  title = {A Revised Radiometric Calibration for the {Hinode}/{EIS} Instrument},
  journal = {A\&A}, volume = 555, pages = {A47}, doi = {10.1051/0004-6361/201220810}
}

@ARTICLE{dzww2025,
  author = {{Del Zanna}, G. and {Weberg}, M.~J. and {Warren}, H.~P.}, year = 2025,
  title = {{Hinode} {EIS}: Updated In-flight Radiometric Calibration},
  journal = {ApJS}, volume = 276, pages = {42}, doi = {10.3847/1538-4365/ad981f}
}

@ARTICLE{weberg2023,
  author = {{Weberg}, M.~J. and {Warren}, H.~P. and {Crump}, N. and {Barnes}, W.}, year = 2023,
  title = {{EISPAC} -- The {EIS} Python Analysis Code},
  journal = {J. Open Source Softw.}, volume = 8, number = 85, pages = {4914}, doi = {10.21105/joss.04914}
}

@ARTICLE{dulk1985,
  author = {{Dulk}, G.~A.}, year = 1985,
  title = {Radio emission from the sun and stars},
  journal = {ARA\&A}, volume = 23, pages = {169--224},
  doi = {10.1146/annurev.aa.23.090185.001125}
}

@ARTICLE{anderson1996,
 author = {{Anderson}, S. W. and {Raymond}, J. C. and {van Ballegooijen}, A. A.},
 year = 1996,
 title = {Ultraviolet Emission-Line Intensities and Coronal Heating by Velocity Filtration: Collisionless Results},
 journal = {ApJ}, volume = 457, pages = {939--948},
 url = {https://articles.adsabs.harvard.edu/pdf/1996ApJ...457..939A}
}

@ARTICLE{doyle2012,
 author = {{Doyle}, J. G. and {Giunta}, A. and {Singh}, A. and {Madjarska}, M. S. and {Summers}, H. and {Kellett}, B. J. and {O'Mullane}, M.},
 year = 2012,
 title = {The Diagnostic Potential of Transition Region Lines Undergoing Transient Ionization in Dynamic Events},
 journal = {Sol. Phys.}, volume = 280, pages = {111--124},
 doi = {10.1007/s11207-012-0025-6}
}

@ARTICLE{viall2012,
 author = {{Viall}, N. M. and {Klimchuk}, J. A.},
 year = 2012,
 title = {Evidence for Widespread Cooling in an Active Region Observed with the {SDO} Atmospheric Imaging Assembly},
 journal = {ApJ}, volume = 753, pages = {35},
 doi = {10.1088/0004-637X/753/1/35}
}

@ARTICLE{testa2026,
 author = {{Testa}, P. and {De Pontieu}, B. and {Cho}, K. and {Ayres}, T. and {Russell}, A. J. B.},
 year = 2026,
 title = {Evidence for Anisotropic Non-{Maxwellian} Velocity Distributions in the Solar Transition Region},
 journal = {ApJL}, volume = 1009, pages = {L19},
 doi = {10.3847/2041-8213/aea155}
}

@ARTICLE{maccormack2017,
  author = {{Mac Cormack}, C. and {V\'{a}squez}, A.~M. and
            {L\'{o}pez Fuentes}, M. and {Nuevo}, F.~A. and
            {Landi}, E. and {Frazin}, R.~A.}, year = 2017,
  title = {Energy Input Flux in the Global Quiet-Sun Corona},
  journal = {ApJ}, volume = 843, pages = {70},
  doi = {10.3847/1538-4357/aa76e9}
}

@ARTICLE{fontenla1991,
  author = {{Fontenla}, J.~M. and {Avrett}, E.~H. and {Loeser}, R.}, year = 1991,
  title = {Energy Balance in the Solar Transition Region. II. Effects of Pressure and Energy Input on Hydrostatic Models},
  journal = {ApJ}, volume = 377, pages = {712--725},
  doi = {10.1086/170399}
}

@ARTICLE{cranmerwinebarger2019,
  author = {{Cranmer}, S.~R. and {Winebarger}, A.~R.}, year = 2019,
  title = {The Properties of the Solar Corona and Its Connection to the Solar Wind},
  journal = {Annu. Rev. Astron. Astrophys.}, volume = 57, pages = {157--187},
  doi = {10.1146/annurev-astro-091918-104416}
}

@ARTICLE{coburn2026,
  author = {{Coburn}, J.~T. and {Innocenti}, M.~E. and {Verscharen}, D.
    and {Artemyev}, A. and {Halekas}, J. and {Jiang}, W. and {Kunz}, M.~W.
    and {Pueschel}, M.~J. and {Riquelme}, M.
    and {\v{S}tver\'{a}k}, \v{S}. and {Wilson III}, L.~B.}, year = 2026,
  title = {Electron-Mediated Transport in Space, Astrophysical, and Laboratory Plasmas},
  journal = {Space Sci. Rev.}, volume = 222, pages = {75},
  doi = {10.1007/s11214-026-01328-0}
}
\endgroup
\end{document}